# Knowledge Extraction for Discriminating Male and Female in Logical Reasoning from Student Model

A. E. E. ElAlfi
Dept. of Computer Science
Mansoura University
Mansoura Egypt, 35516
.

M. E. ElAlami
Dept. of Computer Science
Mansoura University
Mansoura Egypt, 35516

Y. M . Asem
Dept. of Computer Science
Taif University
Taif, Saudia Arabia

*Abstract*:-

The learning process is a process of communication and interaction between the teacher and his students on one side and between the students and each others on the other side. Interaction of the teacher with his students has a great importance in the process of learning and education. The pattern and style of this interaction is determined by the educational situation, trends and concerns, and educational characteristics.

Classroom interaction has an importance and a big role in increasing the efficiency of the learning process and raising the achievement levels of students. Students need to learn skills and habits of study, especially at the university level. The effectiveness of learning is affected by several factors that include the prevailing patterns of interactive behavior in the classroom. These patterns are reflected in the activities of teacher and learners during the learning process. The effectiveness of learning is also influenced by the cognitive and non cognitive characteristics of teacher that help him to succeed, the characteristics of learners, teaching subject, and the teaching methods.

This paper presents a machine learning algorithm for extracting knowledge from student model. The proposed algorithm utilizes the inherent characteristic of genetic algorithm and neural network for extracting comprehensible rules from the student database. The knowledge is used for discriminating male and female levels in logical reasoning as a part of an expert system course.

## I. INTRODUCTION

The learning environment is one of the major task variables that has a special concern from researchers for a long time, in order to identify the factors that may affect its efficiency. The process of interaction within the classroom has a large share of their studies, and they have concluded that the classroom interaction is the essence of the learning process[1].

The classroom interaction which is represented by the communication patterns between the parties of the education and learning process plays an important role in the learners performance, their achievements and their behavioral patterns. Therefore, it is the way to the establishment of ties of understanding between teacher and learners and between learners themselves, and it is the facilitator to understand the goals of education strategies and how to achieve them [2].

The learning skills are indispensable to every student in any area of science. They are inherent in the learner because of its significant impact on his level of collections. This level depends on the quality of the used manner or method in the learning process [3]. Learning skills allow the learner to acquire patterns of behavior that will be associated with him during the course of study. These patterns become study habits and will have a relative stability adjective with respect to the learner[4].

Students in the university have the responsibility to identify their goals and pursue strategies that lead to the achievement of these objectives. Therefore, these strategies should include the study habits, which lead to develop the composition of the student's knowledge [5].

The importance of following good habits of study, which result in reducing students' level of concern for their examination, the high level of self-confidence, and the development of positive attitudes towards the faculty members and the materials was presented by [6]. As a result, the students' achievement will increase as well as their self-satisfaction also [7].

Motivation is also of great importance in raising the tendency towards individual learning. It is one of the basic conditions which achieve the goal of the learning process, the learning ways of thinking, the formation of attitudes and values, the collection of information and the problem solving [8].

The achievement motivation is one of the main factors that may be linked to the objectives of the school system. The students assistance to achieve this motivation will lead to revitalize the level of performance and motivation in order to achieve the most important aspects of school work [9] .

Logical Reasoning lets the individuals think logically to solve the problems, which proves the logical ability of each individual. Induction or inductive reasoning, sometimes called inductive logic, is reasoning which takes us beyond the confines of our current evidence or knowledge to conclusions about the unknown [10].

If the variables of the classroom interaction, the learning and studying skills and the motivation are the whole factors affecting learning, is it possible to compensate each other?

The current study aims to:

1. Identify the differences between female and male in logical reasoning, learning skills, achievement motivation, and their understanding to the efficiency of classroom interaction.

2. Determine the relation between the learning skills, the achievement motivation and the logical reasoning.

3. Present a method for knowledge extraction from the student module in e-learning system.

## II. PROBLEM AND OBJECTIVS OF STUDY

Most researchers agree that the classroom interaction is the essence of the quality of teaching process, and its results are often positive. Also, the pattern and quality of this interaction not only determine the learning situation but also the trends, the concerns, and some aspects of the students' personality.

In Saudi universities, the educational environment of male and female are different. Male students have successful interaction, because the teacher is allowed to observe students and what they do in the classroom. In female environment it is not permissible to watch what happened in the classroom. Logically, this difference may be considered as an advantage to male students. However, female students achievements showed superiority than the male students. This prompted the following questions:

Are there other intermediate variables among the learning environment , the classroom interaction and the student achievement؟. Do these variables affect the student achievement and compensate the classroom interaction؟. Can we extract knowledge by data mining from student model.؟

Accordingly, the problem of the current study determines the following hypotheses:

1. There are statically differences between the female and male students in logical reasoning in faculty of information and computer science at Taif university, Saudi Arabia .

2. There are statically differences between the female and male degrees in learning skills.

3. There are statically differences between the female and male degrees in achievement motivation.

4. There are statically differences between the female and male degrees in understanding the efficiency of classroom interaction.

5. There are statically differences between the female and male degrees in learning skills and logical reasoning when the efficiency of classroom interaction is fixed.

6. There are statically differences between the female and male degrees in achievement motivation and logical reasoning when the efficiency of classroom interaction is fixed.

Then, the study discusses the following questions:

1. Can we provide a machine learning algorithm to extract useful knowledge from the available students data?

2. Can the extracted knowledge from the students data discriminate between the male and female students in the logical reasoning score?

## III. EFFECTIVE STUDENT ATTRIBUTES

The student model plays an important role in the process of teaching and learning. If the elements of this model are chosen properly we can get an important students database. This database can provide useful knowledge when using data mining techniques. Learning skills, achievement motivation, classroom interaction and logical reasoning are the main effective dimensions in student model presented in this study. The following section explains these features.

*A. Learning skills*

A set of behaviors or practices used by the learner during studying the school material. It is determined by the degree which the student obtained through the measure used in the present study

*B. Achievement motivation*

Achievement motivation was looked at as a personality trait that distinguished persons based on their tendency or aspiration to do things well and compete against a standard of excellence [11].

Motivation is the internal condition that activates behavior and gives it direction; energizes and directs goal-oriented behavior. Motivation in education can have several effects on how students learn and how they behave towards subject matter [12]. It is composed of several internal and external motives that affect the behavior of students, orientation and activate individual in different positions to achieve excellence.

## C. Classroom Interaction

During classroom lessons, teachers promote, direct, and mediate discussions to develop learners' understandings of the lesson topics. In addition, teachers need to assess the learners' understanding of the material covered, monitor participation and record progress. Discussions in classrooms are frequently dominated by the more outgoing learners, while others, who may not understand, are silent. The teacher is typically only able to obtain feedback from a few learners before determining the next course of action, possibly resulting in an incomplete assessment of the learners' grasp of the concepts involved. Learners who are not engaged by the discussion, are not forming opinions or constructing understanding, and may not be learning. Classroom distractions can become the focus of attention unless there is some compelling reason for the learner to participate in the discussion [13].

## D. Logical Reasoning

Reasoning is the process of using existing knowledge to draw conclusions, make predictions, or construct explanations. Three methods of reasoning are the deductive, inductive, and abductive approaches. Deductive reasoning starts with the assertion of a general rule and proceeds from there to a guaranteed specific conclusion. Inductive reasoning begins with observations that are specific and limited in scope, and proceeds to a generalized conclusion that is likely, but not certain, in light of accumulated evidence. One could say that inductive reasoning moves from the specific to the general. Abductive reasoning typically begins with an incomplete set of observations and proceeds to the likeliest possible explanation for the set [10].

## IV. APPLICATIONS AND RESULTS

### A. Sample of Study

With regard to the population of students participating in the experiment was 95, (47 of them female and 48 male). These students have studied an expert system course using CLIPS language [14].

### B. Tools of Study

Three measures have been prepared; learning skills, achievement motivation and classroom interaction.

1. Learning skill

A set of 47 clauses reflect the learning skills presented to the students during their study. These clauses dealt with 7 skills. The skills are; management of dispersants, management of the study time, summing and taking notes, preparing for examinations, organization of information, continuation of study, the use of computer and Internet. The student has to choose one of three alternatives (always, sometimes, or never). Their evaluations are 3, 2, or 1. Psychometric measures of the indicator were calculated as follows;

- Criteria Validity

This measure was applied on student sample of 40 male and female students in the faculty of computers and information systems, Taif University, Saudi Arabia. The correlation between their total degrees was 0.82. which is statistically significant at 0.01 level. So, it indicates the validity of the measure.

- Internal Consistency Validity

The correlations between each item and its indicator were calculated. The correlation values vary between 0.37 and 0.65 which are significant at the levels 0.01 and 0.05. Also, the correlation between the total degree and the degrees of each measure are calculated as showin in table I.

TABLE I. THE CORRELATION COEFFICIENT VALUES

| Learning skills | Correlation coefficient | Significant level |
|---|---|---|
| Management of dispersants | 0.76 | |
| Management of the study time | 0.70 | |
| Summing and taking notes | 0.81 | |
| Preparing for examinations | 0.69 | 0.05 |
| Organization of information | 0.82 | |
| Continuation of study | 0.88 | |
| The use of computer & Internet | 0.79 | |

- Indicator reliability

The indicator reliability was measured by two methods as shown in table II.

TABLE II. THE RELIABILITY VALUES OF LEARNING SKILLS MEASURE

| Learning skills | Re-application | | Cronbach's α |
|---|---|---|---|
| | Correlation coefficient | Significant level | |
| Management of dispersants | 0.77 | | 0.75 |
| Management of the study time | 0.66 | | 0.67 |
| Summing and taking notes | 0.61 | 0.01 | 0.62 |
| Preparing for examinations | 0.71 | | 0.70 |
| Organization of information | 0.77 | | 0.75 |
| Continuation of study | 0.68 | | 0.69 |
| The use of computer & Internet | 0.81 | | 0.79 |

This table shows high reliability values of the learning skills measure.

2. Achievement motivation

A set of 71 clauses which reflect the achievement motivation were classified into internal and external pivots. The internal achievement motivation includes; challenge, desire to work, ambition, and self-reliance. The external achievement motivation includes; fear of failure, social motivations, awareness of time importance, and competition. Psychometric measures of the indicator were calculated as follows; :

- Criteria Validity

This measure was applied on the same student sample (40 male and female students) in the same faculty. The correlation between their total degrees was 0.79. which is statistically significant at 0.01 level. So, it indicates the validity of the measure .

- Internal Consistency Validity

The correlations between each item and its indicator were calculated. The correlation values varies between 0.37 and 0.74 which are significant at the levels 0.01 and 0.05. Table 3 shows the calculated correlation coefficients.

TABLE III. CORRELATION COEFFICIENT AND THEIR SIGNIFICANT LEVEL

| Achievement motivation | The indicator | Correlation coefficient | Significant level |
|---|---|---|---|
| Internal achievement motivation | Challenge | 0.68 | |
| | Desire to work | 0.69 | |
| | Ambition | 0.66 | |
| | Self-reliance | 0.56 | 0.01 |
| External achievement motivation | Fear of failure | 0.58 | |
| | Social motivations | 0.71 | |
| | Awareness of time importance | 0.64 | |
| | Competition | 0.56 | |

- Indicator reliability

The indicator reliability was measured by two methods as shown in table 4.

TABLE IV. THE RELIABILITY VALUES OF ACHIEVEMENT MOTIVATION MEASURE

| Learning skills | Correlation coefficient ρ | Significant level | Cronbach's α |
|---|---|---|---|
| challenge | 0.74 | | 0.71 |
| Desire to work | 0.81 | | 0.78 |
| ambition | 0.71 | | 0.72 |
| self-reliance | 0.66 | 0.01 | 0.65 |
| fear of failure | 0.73 | | 0.70 |
| social motivations | 0.68 | | 0.66 |
| awareness of time importance | 0.62 | | 0.64 |
| competition | 0.59 | | 0.61 |

This table shows high reliability values of the achievement motivation measure.

3. Classroom interaction

A set of 27 clauses which measure the level of the classroom interaction were prepared.

- Criteria Validity

This measure was applied on the same student sample (90 male and female students). The principle components method is used for factor analysis.

Getman criterion for factor analysis was used to determine the number of factors. Varimax orthogonal rotation was also used. These two methods yield to the extraction of three factors (saturation ≥ ± 3 ). Each new factor has ≥ three factors. Table 5 shows the results of the factor analysis.

TABLE V. THE RESULTS OF THE FACTOR ANALYSIS.

| Clause No. | Factor | | | Clause No. | Factor | | |
|---|---|---|---|---|---|---|---|
| | First | Second | Third | | First | Second | Third |
| 1 | 0.45 | | | 14 | 0.35 | | |
| 2 | | 0.51 | | 15 | 0.40 | | |
| 3 | 0.41 | | | 16 | 0.51 | | |
| 4 | | 0.44 | | 17 | 0.38 | | |
| 5 | 0.61 | | | 18 | 0.47 | | |
| 6 | | 0.55 | | 19 | 0.46 | | |
| 7 | | 0.52 | | 20 | 0.45 | | |
| 8 | | 0.63 | | 21 | | 0.44 | |
| 9 | 0.66 | | | 22 | | 0.51 | |
| 10 | | 0.46 | | 23 | | | 0.61 |
| 11 | 0.44 | | | 24 | | 0.39 | |
| 12 | | 0.55 | | 25 | | | 0.55 |
| 13 | 0.63 | | | 26 | | | 0.59 |
| Eigen values | | | | | 3.84 | 3.8 | 2.32 |
| Variance | | | | | 14.23 | 14.08 | 8.59 |

This table shows that the measure has saturated by 3 factors:

The first factor is saturated with 13 individual items. These items revolve around the lecturer's ability to manage the classroom interaction. This factor may be defined as teacher's positivity.

The second factor has a saturation of 10 items that revolve around the student's ability to interact with the lecturer on the basis of the lecture theme. This factor may be defined as student's positivity.

The third factor has a saturation of 3 items only. It revolves around the potential of the classroom that facilitate the process of interaction between the student and lecturer. This might be called the potential of the classroom. The factor analysis has deleted the factor number 27.

- Internal Consistency Validity

The correlations between each item and its indicator were calculated. Table 6 shows the calculated correlation coefficients. This table indicates that the individual factors are correlated to their main factors (1st, 2nd and 3rd) which proves internal consistency of the measure.

TABLE VI. THE CORRELATION COEFFICIENTS

| 1st Factor | | | 2nd Factor | | | 3rd Factor | | |
|---|---|---|---|---|---|---|---|---|
| Fact. No. | ρ | Sign. level | Fact. No. | ρ | Sign. level | Fact. No. | ρ | Sign. level |
| 1 | 0.49 | 0.01 | 2 | 0.45 | 0.01 | 23 | 0.45 | 0.01 |
| 3 | 0.61 | | 4 | 0.46 | | 25 | 0.36 | |
| 5 | 0.39 | | 6 | 0.44 | | 26 | 0.35 | |
| 9 | 0.38 | | 7 | 0.52 | | | | |
| 11 | 0.45 | | 8 | 0.51 | | | | |
| 13 | 0.42 | | 10 | 0.59 | | | | |
| 14 | 0.35 | | 12 | 0.43 | | | | |
| 15 | 0.61 | | 21 | 0.51 | | | | |
| 16 | 0.52 | | 22 | 0.60 | | | | |
| 17 | 0.46 | | 24 | 0.42 | | | | |
| 18 | 0.59 | | | | | | | |
| 19 | 0.52 | | | | | | | |
| 20 | 0.42 | | | | | | | |

- Indicator reliability

The indicator reliability was measured by two methods as shown in table 7.

TABLE VII. THE RELIABILITY VALUES OF THE CLASSROOM INTERACTION MEASURE

| Learning skills | Correlation coefficient ρ | Significant level | Cronbach's α |
|---|---|---|---|
| Teacher's positivity | 0.79 | 0.01 | 0.80 |
| Student's positivity | 0.77 | | 0.75 |
| Potential of the classroom | 0.69 | | 0.68 |

So, the above table shows that the measure of the classroom interaction has an acceptable degree of consistency.

C. *Testing the study hypotheses*

- The first hypothesis

There are statistical differences between the mean scores of the female and male students in logical reasoning in the faculty of information and computer science at Taif university, , Saudi Arabia . To verify this hypothesis the t test was used to measure the differences between the means of the independent groups. The results are shown in the following table.

TABLE VIII. THE T TEST VALUE FOR THE DIFFERENCES OF MALE AND FEMALE STUDENTS IN LOGICAL REASONING

| Gender | Number | Mean | Standard deviation | T | Significant level |
|---|---|---|---|---|---|
| Male | 49 | 11.84 | 2.86 | 3.99 | 0.01 |
| Female | 48 | 13.73 | 1.67 | | |

The above table shows that there are statistical differences between the mean scores of the males and females in the logical reasoning in favor of females. Also, this result indicates the superiority of females in logical reasoning ability to understand the linkage of precondition and conclusion .

- The second hypothesis

There exist mean differences between female and male degrees in learning skills. To verify this hypothesis the analysis of variance of the multi-variables (MANOVA) was used. Both the Box test for homogeneity of the matrix and the value of the Levene test of equal contrast were insignificant for all dimensions. Wilks Lambda test value is equal to 0.68 which is significant. The ETA value is equal to 0.32. These results indicate the validity of the test and give an indication of the existence of differences in accordance with the type of learning skills. The following table shows the results of the analysis of variance test.

TABLE IX. THE ANALYSIS OF VARIANCE OF THE MULTI-VARIABLES (MANOVA) IN LEARNING SKILLS

| | Dimensions | Sum of squares | Degree of freedom | Mean square | F | Significant level | η |
|---|---|---|---|---|---|---|---|
| Type | Management of dispersants | 18.05 | 1 | 18.045 | 1.46 | Insignificant | 0.02 |
| | Management of the study time | 114.19 | 1 | 114.186 | 16.2 | 0.01 | 0.15 |
| | Summing and taking notes, | 106.23 | 1 | 106.231 | 23.2 | | 0.2 |
| | Preparing for examinations | 20.77 | 1 | 20.716 | 9.88 | | 0.1 |
| | Organization of information | 43.08 | 1 | 43.079 | 12.9 | | 0.12 |
| | Continuation of study | 36.03 | 1 | 36.029 | 10.5 | | 0.10 |
| | The use of computer & Internet | 177.97 | 1 | 177.968 | 15.5 | | 0.14 |
| | Total | 3102.3 | 1 | 3102.31 | 24.5 | | 0.21 |
| Error | Management of dispersants | 1177.3 | 95 | 12.393 | | | |
| | Management of the study time | 669.36 | 95 | 7.046 | | | |
| | Summing and taking notes | 436.02 | 95 | 4.590 | | | |
| | Preparing for examinations | 199.30 | 95 | 2.098 | | | |
| | Organization of information | 317.29 | 95 | 3.340 | | | |
| | Continuation of study | 325.72 | 95 | 3.429 | | | |
| | The use of computer & Internet | 1088.9 | 95 | 11.462 | | | |
| | Total | 12041.8 | 95 | 126.76 | | | |

The above table shows that there are statistical of differences between males and females in the learning skills in all dimensions except the first dimension (Management of dispersants). To measure the differences, the mean and standard deviation were calculated as shown in table 10.

This table shows that there are statistical differences in the learning skills in favor of females. Females are more likely to use the correct methods of learning, more able to manage time and planning to take advantage of it. They are more able to take of observations, notes and summaries. They are more able to prepare well for exams throughout the semester and organize information and use them correctly more than males. Also, they do not delay studying till the end of the year. They use computer and Web to get and exchange information. Females are better in general. The results did not show differences between males and females in management of dispersants, everyone is making effort to overcome them but what is important is what happens after that.

TABLE X. THE MEAN AND STANDARD DEVIATION OF DEGREES IN LEARNING SKILLS

| Dimensions | Gender | Number | Mean | Standard deviation |
|---|---|---|---|---|
| Management of dispersants | M | 49 | 25.408 | 0.503 |
| | F | 48 | 26.271 | 0.508 |
| Management of study time | M | 49 | 15.163 | 0.379 |
| | F | 48 | 17.333 | 0.383 |
| Summing and taking notes | M | 49 | 14.469 | 0.306 |
| | F | 48 | 16.563 | 0.309 |
| Preparing for examinations | M | 49 | 10.367 | 0.207 |
| | F | 48 | 11.292 | 0.209 |
| Organization of information | M | 49 | 10.980 | 0.261 |
| | F | 48 | 12.313 | 0.264 |
| Continuation of study | M | 49 | 9.510 | 0.265 |
| | F | 48 | 10.729 | 0.267 |
| Use of computer & Internet | M | 49 | 12.041 | 0.484 |
| | F | 48 | 14.750 | 0.489 |
| Total | M | 49 | 97.939 | 1.608 |
| | F | 48 | 109.250 | 1.625 |

- The third hypothesis

There are statistical differences between the mean scores of the female and male students in achievement motivation.

To verify this hypothesis the analysis of variance of the multi-variables (MANOVA) test was used. The Box test for homogeneity of the matrix was insignificant. The value of the Levene test of equal contrast, was also insignificant. Wilks Lambda test value is equal to 0.56 which is significant. The value of ETA is 0.44. So, all these results indicate the validity of the test. The following table shows the results of the analysis of variance test and an indicate that the differences are affected by the achievement motivation.

TABLE XI. THE ANALYSIS OF VARIANCE OF THE MULTI-VARIABLES (MANOVA) IN ACHIEVEMENT MOTIVATION.

| | Dimensions | Sum of squares | Degrees of freedom | Mean square | F | Significant level | η |
|---|---|---|---|---|---|---|---|
| Type | challenge | 184.22 | 1 | 184.22 | 24.89 | 0.01 | 0.21 |
| | Desire to work | 20.28 | 1 | 20.28 | 1.46 | not | 0.02 |
| | ambition | 57.63 | 1 | 57.63 | 13.1 | 0.01 | 0.12 |
| | self-reliance | 19.22 | 1 | 19.22 | 4.85 | 0.05 | 0.05 |
| | fear of failure | 14.33 | 1 | 14.33 | 1.65 | not | 0.02 |
| | social motivations | 177.97 | 1 | 177.97 | 16.5 | 0.01 | 0.15 |
| | awareness of time importance | 155.23 | 1 | 155.23 | 29.79 | 0.01 | 0.24 |
| | competition | 7.49 | 1 | 7.49 | 0.59 | not | 0.06 |
| | Total | 3890.4 | 1 | 3960.4 | 12.37 | 0.01 | 0.12 |
| Error | challenge | 703.22 | 95 | 7.4 | | | |
| | Desire to work | 1321.8 | 95 | 13.91 | | | |
| | ambition | 419.22 | 95 | 4.42 | | | |
| | self-reliance | 376.8 | 95 | 3.97 | | | |
| | fear of failure | 825.18 | 95 | 8.69 | | | |
| | social motivations | 1024.92 | 95 | 10.79 | | | |
| | awareness of time importance | 495.1 | 95 | 5.21 | | | |
| | competition | 1197.26 | 95 | 12.6 | | | |
| | Total | 29880.6 | 95 | 314.53 | | | |

The above table shows that there are statistical differences between males and females in achievement motivation in five dimensions. To measure the differences, the mean and standard deviation were calculated as shown in the following table.

TABLE XII. THE MEAN AND STANDARD DEVIATION OF DEGREES IN ACHIEVEMENT MOTIVATION

| Dimensions | Gender | Number | Mean | Standard deviation |
|---|---|---|---|---|
| Challenge | M | 49 | 21.306 | 0.389 |
| | F | 48 | 24.063 | 0.393 |
| Desire to work | M | 49 | 24.898 | 0.533 |
| | F | 48 | 25.812 | 0.538 |
| Ambition | M | 49 | 13.000 | 0.300 |
| | F | 48 | 14.542 | 0.303 |
| Self-reliance | M | 49 | 12.735 | 0.285 |
| | F | 48 | 13.625 | 0.287 |
| Fear of failure | M | 49 | 17.898 | 0.421 |
| | F | 48 | 18.667 | 0.425 |
| Social motivations | M | 49 | 21.041 | 0.469 |
| | F | 48 | 23.750 | 0.474 |
| Awareness of time importance | M | 49 | 18.449 | 0.326 |
| | F | 48 | 20.979 | 0.330 |
| Competition | M | 49 | 21.673 | 0.507 |
| | F | 48 | 22.229 | 0.512 |
| Total | M | 49 | 151.000 | 2.534 |
| | F | 48 | 163.667 | 2.560 |

The above table shows that there are statistical differences in achievement motivation in favor of females in all dimensions except the dimensions of the desire to work, the fear of failure, and competition. This means that females have external incentives which lead them to exert effort, such as the motives of the desire to challenge the male society significantly, as if to prove a kind of self-motivation, ambition and self-reliance. Also, it seems that they were in need to change their society's perception that they must rely only on men in everything, and they are motivated by external motivation like satisfaction of parents, acquiring others' admiration and attract their attention, awareness of the importance of time, and they achieve success in running time.

- The fourth hypothesis

There are statistical differences between the female and male degrees in understanding the efficiency of classroom interaction.

To verify this hypothesis the analysis of variance of the multi-variables (MANOVA) test was used. The Box test for homogeneity of the matrix was insignificant. The value of the Levene test of equal contrast, was also insignificant. Wilks Lambda test value is equal to 0.82 which is significant. The value of ETA is 0.18. All these results indicate the validity of the test and prove that the differences are affected by the type of classroom interaction. The following table shows the results of the analysis of variance test.

TABLE XIII.  THE ANALYSIS OF VARIANCE OF THE MULTI-VARIABLES (MANOVA) IN CLASSROOM INTERACTION

| | Dimensions | Sum of squares | Degree of freedom | Mean square | F | Significant level | η |
|---|---|---|---|---|---|---|---|
| Type | Potential of the classroom. | 14.937 | 1 | 14.937 | 5.06 | 0.05 | 0.051 |
| | Student's positivity | 11.376 | 1 | 11.376 | 0.88 | not | 0.009 |
| | Teacher's positivity | 127.436 | 1 | 127.44 | 6.57 | 0.05 | 0.065 |
| | Total | 138.786 | 1 | 138.79 | 2.38 | not | 0.024 |
| Error | Potential of the classroom. | 280.692 | 95 | 2.955 | | | |
| | Student's positivity | 1229.08 | 95 | 12.938 | | | |
| | Teacher's positivity | 1842.585 | 95 | 19.396 | | | |
| | Total | 5537.17 | 95 | 58.286 | | | |

The above table shows that the F values are significant in the dimensions of teacher's positivity and the potential of the classroom. To measure the differences, the mean and standard deviation were calculated as shown in the following table.

TABLE XIV.  THE MEAN AND STANDARD DEVIATION OF DEGREES IN CLASSROOM INTERACTION

| Dimensions | Gender | Number | Mean | Standard deviation |
|---|---|---|---|---|
| Potential of the classroom | M | 49 | 5.327 | 0.246 |
| | F | 48 | 4.542 | 0.248 |
| Student's positivity | M | 49 | 22.878 | 0.514 |
| | F | 48 | 23.563 | 0.519 |
| Teacher's positivity | M | 49 | 29.959 | 0.629 |
| | F | 48 | 27.667 | 0.636 |
| Total | M | 49 | 58.163 | 1.091 |
| | F | 48 | 55.771 | 1.102 |

The above table shows that there are statistical differences in the dimensions of potential of the classroom and teacher's positivity in favor of males. This means that males interact better than females in the classroom, particularly in the dimensions of teacher's positivity and potential of the classroom. This can be attributed to the nature of teaching to the male students as there is direct and face to face interaction. However, in the absence of direct interaction, females feel that the learning environment is not valid, the lecturer does not do his utmost in the commentary. Hence, the problem is not the women from their point of view.

- The fifth hypothesis

There are statistical differences between the female and male degrees in learning skills and logical reasoning when the efficiency of classroom interaction is fixed. The partial correlation coefficient between the degrees in learning skills and logical reasoning is used to verify this hypothesis while the efficiency of classroom interaction is fixed. The results are shown in the following table.

TABLE XV.  THE CORRELATION BETWEEN THE LEARNING SKILLS AND LOGICAL REASONING

| Dimensions | Male | | Female | |
|---|---|---|---|---|
| | Correlation coefficient ρ | Significant level | Correlation coefficient ρ | Significant level |
| Management of dispersants | 0.7 | 0.01 | 0.64 | 0.01 |
| Management of study time | 0.45 | 0.01 | 0.4 | 0.01 |
| Summing and taking notes | 0.26 | Not | 0.35 | 0.01 |
| Preparing for examinations | 0.23 | Not | 0.24 | Not |
| Organization of information | 0.35 | 0.05 | 0.39 | 0.01 |
| Continuation of study | 0.3 | 0.05 | 0.45 | 0.01 |
| The use of computer & Internet | 0.27 | Not | 0.31 | 0.05 |
| Total | 0.57 | 0.01 | 0.64 | 0.01 |

The above table shows the following:

For males: There is positive correlation coefficient between the degree of learning skills and levels of logical reasoning in the dimensions of management of dispersants, Management of the study time, the organization of information, Continuation of study, and the total degree. This means that , if the student is more able to focus, manage time, organize information, and study continuously without delay, it is expected to achieve highly in the degree of logical reasoning. This result agrees with the nature of the material needs to get a high degree of focus, organization and effort unlike any other material.

As for females: there is a correlation coefficient between the degree of learning skills and the levels of logical reasoning in all dimensions except preparing for examinations. Correlation has appeared in the dimensions of summing and taking notes, the use of computers and the Web to access information. Consequently, the logical reasoning degrees are affected by the same factors as in the male case in addition to the latter two dimensions.

- The sixth hypothesis

There are statistical differences between female and male degrees in achievement motivation and logical reasoning when the efficiency of classroom interaction is fixed.

To verify this hypothesis the partial correlation coefficient test between the degrees of achievement motivation and logical reasoning while the classroom interaction is fixed was used. This was done due to the presence of differences among the achievement motivation, logical reasoning and understanding of the efficiency of classroom interaction. The results are shown in the following table.

TABLE XVI. THE CORRELATION COEFFICIENT VALUES AND THE SIGNIFICANCE BETWEEN ACHIEVEMENT MOTIVATION AND LOGICAL REASONING.

| Dimensions | Male | | Female | |
|---|---|---|---|---|
| | Correlation coefficient ρ | Significant level | Correlation coefficient ρ | Significant level |
| Challenge | 0.52 | | 0.66 | |
| Desire to work | 0.54 | | 0.63 | |
| Ambition | 0.40 | | 0.52 | |
| Self-reliance | 0.41 | | 0.75 | |
| Fear of failure | 0.44 | 0.01 | 0.55 | 0.01 |
| Social motivations | 0.53 | | 0.60 | |
| Awareness of time importance | 0.47 | | 0.63 | |
| Competition | 0.56 | | 0.61 | |
| Total | 0.61 | | 0.79 | |

The above table shows a correlation between the degree of achievement motivation and levels of logical reasoning for both male and female. This is very important, since the nature of the course requires large student's motivation to deal with.

They need to exert an effort, regardless of what is behind this effort, and this result agrees with the majority of studies that proved a positive relationship between achievement motivation and achievement.

*D. Knowledge Extraction*

This section well illustrate the students database description in addition to discussing the study questions.

- Students database description

A student model database used for knowledge extraction is composed of four main predictive measures and one target measure. The first measure is the learning skills which includes 7 attributes namely: management of dispersants, management of the study time, summing and taking notes, preparing for examinations, organization of information, continuation of study, and the use of computer & internet. The second measure is achievement motivation which is divided into internal and external motivations. Internal motivations includes 4 attributes; challenge, the desire to work, ambition, and self-reliance. External motivations includes 4 attributes; fear of failure, social motivations, awareness of time importance, and competition. The third measure is classroom interaction which includes potential of the classroom, student's positivity, and teacher's positivity. The final measure is student score in the expert system course which is divided into 5 test units. In deed, the target measure is logical reasoning .

- The first question:

Can we provide a machine learning algorithm to extract useful knowledge from the available students data ?

Data mining (DM) or in other words "the extraction of hidden predictive information from data" is a powerful new technology with great potential to help users focus on the most important information in large data sets. The general goal of DM is to discover knowledge that is not only correct, but also comprehensible and interesting for the user. Among the various DM algorithms, such as clustering, association rule finding, data generalization and summarization, classification is gaining significant attention [ 15].

Classification is the process of finding a set of models or functions which describe and distinguish data classes or concepts, for the purpose of being able to use the model to predict the class of objects whose class label is unknown. In classification, a rule generally represents discovered knowledge in the form of IF-THEN rules. The classification methods can be categorized into two groups, non-rule-based and rule-based classification groups [16]. Non-rule-based classification methods are such as artificial neural network (ANN) [17-18] and support vector machines [19]. Rule-based classification methods are such as C4.5 [20], and decision table [21]. Rule-based classification methods directly extract hidden knowledge from the data. However, non-rule-based classification methods are generally more accurate than rule-based classification methods.

This section presents the proposed algorithm for extracting a set of accurate and comprehensible rules from the input database via trained ANN using genetic algorithm (GA). The details of the proposed algorithms is explained in previous work [22]. A concise algorithm for extracting a set of accurate rules is shown in the following steps:

1. Assume that;

   1.1 The input database has N predictive attributes plus one target attribute.

   1.2 Each predictive attribute has a number of values, and can be encoded into binary sub-string of fixed length.

   1.3 Each element of a binary sub-string equals one if its corresponding attribute value exists, while all the other elements are equal to zero.

   1.4 Repeat the steps (1.2) and (1.3) for each predictive attribute, in order to construct the encoded input attributes' vectors.

   1.5 The target attribute has a number of different classes, and can be encoded as a bit vector of a fixed length as explained in step (1.3).

2. The ANN is trained on the encoded vectors of the input attributes and the corresponding vectors of the output classes until the convergence rate between the actual and the desired output will be achieved.

3. The exponential function of each output node of ANN can be constructed as a function of the values of the input attributes and the extracted weights between the layers.

4. To find the rule belongs to a certain class, GA is used to find the optimal chromosome (values of input attributes), which maximizes the output function of the corresponding node (class) of the ANN.

5. The extracted chromosome must be decoded to find the corresponding rule as follows;

   5.1 The optimal chromosome is divided into N segments.

   5.2 Each segment represents one attribute, and has a corresponding bits length represent their values.

   5.3 The attribute values are existed if the corresponding bits in the optimal chromosome are equal to one and vice versa.

   5.4 The operators "OR" and "AND" are used to correlate the existing values of the same attribute and the different attributes, respectively.

   5.5 The extracted rules must be refined to cancel the redundant attributes.

- The second question:

Can the extracted knowledge from the students data discriminate between the male and female students in the logical reasoning score ?

This question will be dealt with through the following rules extraction and their interpretations.

Assume the following abbreviations:

F : Fail, P : Pass, G : Good, V.G : Very Good, L : Low, M : Medium, H : High, Ma : Male, Fe : Female.

1. If Unit 1 = F → Then Reasoning = F.
2. If Unit 2 = F → Then Reasoning = F.
3. If Unit 5 = F → Then Reasoning = F.
4. If Unit 1= V.G or Unit 2 = V.G and Maintaining learning = H → Then Reasoning = V.G.
5. If Unit 1 = V.G or Unit 2 = V.G and Fear of failure = H → Then Reasoning = V.G.

From the above rules one can conclude that units number 1, 2, and 5 are the most effective attributes in the final results. This is because they include the principles, the inductive reasoning and the object oriented programming in CLIPS respectively.

6. If Unit 1 = P and Unit 3 = F or Unit 4 = F → Then Reasoning = P.
7. If Ambition = H and Unit 3 = F or Unit 4 = F → Then Reasoning = P.
8. If Self-reliance = M and Unit 3 or Unit 4 = F → Then Reasoning = P.

The rules numbers 6, 7, and 8 indicate that units number 3 and 4 are not effective. The high ambition and medium self-reliance lead to passing in reasoning although the fail score in unit 3 or unit 4.

9. If Gender = Ma and Ambition = L → Then Reasoning = F.
10. If Gender = Fe and Ambition = L → Then Reasoning = P.
11. If Gender = Ma and Management of dispersants = L → Then Reasoning = F.
12. If Gender = Fe and Management of dispersants = L → Then Reasoning = P.
13. If Gender = Ma and Self-reliance = L → Then Reasoning = F.
14. If Gender = Fe and Self-reliance = L → Then Reasoning = P.
15. If Gender = Fe and The desire to work = M and organization of information = H → Then Reasoning = G.
16. If Gender = Fe and Fear of failure = M and Self-reliance = H → Then Reasoning = Good.
17. If Gender = Fe and organization of information = H and Maintaining learning = M → Then Reasoning =G.
18. If Gender = Fe and The potential class = L and Unit 2 = P → Then Reasoning = G.
19. If Gender = Fe and Time management = H and organization of information = H and unit 3 = V.G → Then Reasoning = V.G

The previous rules clarify the attributes effect on the reasoning results taking into consideration the effect of the gender attribute.

V. CONCLUSIONS

It is our intent to explore how data mining is being used in education services at Taif University in Saudi Arabia. Educational data mining is the process of converting raw data from educational systems to useful information that can be used to inform design decisions and answer research questions.

The importance of the study can be stated as follows:

It is dealing with the learning environment of Saudi Arabia, that has a special nature in the education of females and the factors affecting it . The study combines the variables related to personality, mental and environmental aspects in order to reach an integrated view of the learning nature process and the factors affecting it. It addresses the subject of study habits and achievement motivations, which are important issues that affect the educational process. Good study habits will help students in the collection of knowledge, the achievement motivation and push them to the challenge of the obstacles to achieve their goals. The study presents an efficient technique that utilizes artificial neural network and genetic algorithm for extracting comprehensive rules from student database. The extracted knowledge supports the effective attributes that are most effective in the final score of the logical reasoning.

## VI. FUTURE WORKS

E-learning represents a great challenge in education, that large amounts of information are continuously generated and available. Using data mining to extract knowledge from information is the best approach to process the obtained knowledge in order to identify the student needs. Tracking student behavior in virtual e-learning environment makes the web mining of the resulting databases possible, which encourages the educationalists and curricula designers to create learning contents. So, we aim at introducing a novel rule extraction method that depends on Fuzzy Inductive Reasoning methodology. This method has been driven from a data set obtained from a virtual campus e-learning environment. Hence, to gain the best benefit from this knowledge, the results should be described in terms of a set of logical rules that trace the different level of the student performance.

## References

[1] Kudret Ozkal, Ceren Tekkaya, Jale Cakiroglu, Semra Sungur, "A conceptual model of relationships among constructivist learning environment perceptions, epistemological beliefs, and learning approaches", Learning and Individual Differences, Volume 19, Issue 1, 1st Quarter, Pages 71-79, 2009.

[2] Shun Lau, Youyan Nie, "Interplay Between Personal Goals and Classroom Goal Structures in Predicting Student Outcomes: A Multilevel Analysis of Person–Context Interactions", Journal of Educational Psychology, Volume 100, Issue 1, Pages 15-29, February 2008.

[3] Karin Tweddell Levinsen, "Qualifying online teachers- Communicative skills and their impact on e-learning quality", Education and Information Technologies, Volume 12, Number 1 / March 2007.

[4] Richards, L.G, " Further studies of study habits and study skills", Frontiers in Education Conference, 31st Annual, Volume 3, Page(s):S3D - S13, 10-13 Oct. 2001.

[5] Nneji, L . M, "Study habits of Nigerian University Students", Nigerian Educational Research, Development Council , Abuja , Nigeria , Pages 490 – 495, 2002.

[6] Okapala. A, Okapala. C, Ellis. R, "Academic Efforts and study habits among students in a principles of macroeconomics course", Journal of Education for Business , 75 (4), Pages 219 – 224, 2000.

[7] Marcus Credé, Nathan R. Kuncel, " Study Habits, Skills, and Attitudes: The Third Pillar Supporting Collegiate Academic Performance", Perspectives on Psychological Science, Volume 3, Issue 6, Pages: 425-453, November 2008,

[8] Weiqiao Fan, Li-Fang Zhang, " Are achievement motivation and thinking styles related? A visit among Chinese university students", Learning and Individual Differences, Volume 19, Issue 2, Pages 299-303, June 2009.

[9] Ricarda Steinmayr, Birgit Spinath, "The importance of motivation as a predictor of school achievement", Learning and Individual Differences, Volume 19, Issue 1, Pages 80-90, 1st Quarter 2009.

[10] Yuichi Goto, Takahiro Koh, Jingde Cheng, "A General Forward Reasoning Algorithm for Various Logic Systems with Different Formalizations", 12th International Conference, Knowledge-Based Intelligent Information & Engineering Systems, Proceedings Part II, Pages 526-535, September 3-5, 2008.

[11] Wigfield, A, & Eccles, J.S, "Development of achievement motivation", San Diego, San Francisco, New York, Boston, London, Sydney, Tokyo: Academic Press, 2002.

[12] David C, McClelland, "Methods of Measuring Human Motivation", in John W. Atkinson, ed., Motives in Fantasy, Action and Society (Princeton, N.J.: D. Van Nos-trand, Pages 12-13, 1958.

[13] Timothy W. Pelton & Leslee Francis Pelton, "The Classroom Interaction System", (CIS): Neo-Slates for the Classroom" W.-M. Roth (ed.), CONNECTIONS '03, Pages 101–110, 2003.

[14] Joseph C. Giarratano, "CLIPS User's Guide", Version 6.2, March 31st 2002.

[15] Li Liu, Murat Kantarcioglu, Bhavani Thuraisingham, "The applicability of the perturbation based privacy preserving data mining for real-world data", Data & Knowledge Engineering, Volume 65, Issue 1, Pages 5-2, April 2008.

[16] Tan, C., Yu, Q., & Ang, J. H., "A dual-objective evolutionary algorithm for rules extraction in data mining", Computational Optimization and Applications, 34, Pages 273–294, 2006.

[17] Humar Kahramanli and Novruz Allahverdi, "Rule extraction from trained adaptive neural networks using artificial immune systems", Expert Systems with Applications 36, Pages 1513–1522, 2009.

[18] Richi Nayak, "Generating rules with predicates, terms and variables from the pruned neural networks", Neural Networks 22, Pages 405-414, 2009.

[19] J.L. Castro, L.D. Flores-Hidalgo, C.J. Mantas and J.M. Puche, "Extraction of fuzzy rules from support vector machines", Fuzzy Sets and Systems, Volume 158, Issue 18, Pages 2057-2077, 16 September 2007.

[20] Kemal Polat and Salih Güneş, "A novel hybrid intelligent method based on C4.5 decision tree classifier and one-against-all approach for multi-class classification problems", Expert Systems with Applications, Volume 36, Issue 2, Part 1, Pages 1587-1592, March 2009.

[21] Yuhua Qian, Jiye Liang and Chuangyin Dang, "Converse approximation and rule extraction from decision tables in rough set theory", Computers & Mathematics with Applications, Volume 55, Issue 8, Pages 1754-1765, April 2008.

[22] A. Ebrahim ELAlfi, M. Esmail ELAlami, R. Haque, "Extracting Rules From Trained Neural Network Using GA For Managing E- Business", Applied Soft Computing 4, Pages 65-77, 2004.